\documentclass[12pt]{iopart}
\usepackage{graphicx}
\usepackage{subcaption}
\usepackage{hyperref}
\def\dashdotted{\xleaders\hbox to 1em{$- \cdot$}\hfill $-$}
\eqnobysec
\begin{document}
\title{Exact solutions for charged spheres and their stability. I. Perfect Fluids.}

\author{ K. Dev}

\address{Department of Physics and Astronomy, Dickinson College, Carlisle PA. }
\ead{devk@dickinson.edu}

\begin{abstract}
We study exact solutions of the Einstein-Maxwell equations for the interior gravitational field of static spherically symmetric charged compact spheres. The spheres are composed of a perfect fluid  with  a charge distribution that creates  a static radial electric field. The inertial mass density  of the fluid has the form $\rho (r) = \rho_o + \alpha r^2$ ($\rho_o$ and $\alpha$ are constants) and the total charge $q(r)$ within a sphere of radius $r$ has the form $q = \beta r^3$ ($\beta$ is a constant). We evaluate the critical values of $M/R$ for these spheres as a function of $Q/R$ and compare these values with those given by the Andr\'{e}asson formula. 
\end{abstract}

\section{Introduction}
The  Reissner-Nordstr\"{o}m  metric (Reissner \cite{Reissner}, Weyl  \cite{Weyl} and Nordstr\"{o}m   \cite{Nordstrom}) is a solution of the Einstein-Maxwell equations that represents the exterior gravitational field of a spherically symmetric charged body.  In fact,  it is the unique asymptotically flat vacuum solution around any
charged spherically symmetric object, irrespective of how that body may be composed
or how it may evolve in time  (Carter \cite{Carter}, Ruback  \cite{Ruback} and Chru\'{s}ciel  \cite{Chrusciel}). It contains no information about its
source other than its total mass and total charge, and probably most interesting it imposes no constraints on its internal
structure. 

The internal structure  of a charged sphere is established by solving the coupled Einstein-Maxwell equations. A solution of these equations  describes  the gravitational field inside the sphere as a function of the distribution of matter and energy within the sphere. The Einstein-Maxwell equations for a charged static spherically distribution of matter  reduce to a set  of three independent non-linear second order differential equations that connect  the metric coefficients $g_{tt}$ and $g_{rr}$ with the physical quantities that represent the matter content of the sphere.  A description of  the matter content of a charged perfect fluid sphere includes functions that defines its inertial mass density $\rho_{fl}$, its pressure $p$ and the electric charge distribution $q$ or the electromagnetic energy density $\rho_{em}$. 
 Thus, the complete system of equations to be solved consists of three linearly independent equations with five unknowns. This situation gives total freedom in choosing two of the five functions and  then solving for the remaining three. It is therefore hardly surprising that there exists very many exact solutions of the Einstein-Maxwell equations for charged spheres.  A comprehensive  review of the various methods of solving the Einstein-Maxwell equations for charged spheres was done by Ivanov \cite{Ivanov}.

An important question that arises in the study of the structure of  spherically symmetric compact objects  is the following: what is the maximum of  value $M/R$ (the total mass of the body divided its radius) that is allowed before gravitational collapse occurs? It is well known that for neutral perfect fluid spheres that the stability limit is given by the Buchdahl  limit \cite{Buchdahl}:\begin{equation}
\frac{M}{R} \leq \frac{4}{9}  {~~~\rm for ~ uncharged ~perfect~ fluid~spheres}.
\end{equation}
In case of charged compact objects this quantity becomes dependent on the total charge $Q$, since the  addition of charge to the system  increases its total  energy and hence its total mass.   Andr\'{e}asson \cite{Andreasson}, 
has published a remarkable result that claims that for any compact spherically symmetric charged distribution  the following relationship  between $M/R$ and $Q/R$ holds:
\begin{equation}
\label{Andf}
\frac{M}{R} \leq \left(\frac{1}{3} + \sqrt{  \frac{1}{9} + \frac{1}{3} \frac{Q^2}{R^2} } ~\right)^2 {~~~\rm for ~all~charged~ spheres}.
\end{equation}
This formula generalizes the Buchdahl  result for uncharged perfect fluid spheres. It is worth noting that in his derivation, Buchdahl placed the following constraints on the physical properties of the fluid: (i) $d\rho/dr  < 0$ i.e., the density  decreases outward from the center of the sphere  and (ii) the pressure is isotropic.  In the derivation of his formula, Andr\'{e}asson made the following assumptions about  the matter content of the sphere: (i) $ p_r + 2 p_t < \rho$ (in a spherically symmetric  system it is possible to have the radial pressure, $p_r$ different from the tangential pressure, $p_t$) and (ii) $\rho >0$. 
We note that unlike the neutral case    Andr\'{e}asson     did not require the condition $d\rho/dr <0$. 
Numerical investigations  (\cite{Lemos2} and \cite{Lemos}),    of the upper bound of $M/R$ as a function of $Q/R$  have verified that the  Andr\'{e}asson    formula provides a valid upper bound of $M/R$ for charged spheres.   One of our aims in this study is to continue the investigation of   the validity of the  Andr\'{e}asson  limit using a mixture of analytical and numerical techniques.

A class of charge spheres that have received considerable attention in theoretical physics are extremal black holes. Extremal non-rotating charged black holes are interior solutions of the Reissner-Nordstrom metric in which  the mass is equal to the charge in geometric units. These objects have zero surface gravity and their horizon structure is different from that of regular black holes. In general relativity extremal black holes are  important in the study and proof of the third law of black hole thermodynamics. In string theory they are ubiquitous, in particular they were used to derived the Bekenstein-Hawking entropy formula and in general they are easier to describe than regular black holes  quantum mechanically because of their vanishing surface gravity and consequently vanishing temperature for Hawking radiation \cite{Vafa}.  In many solutions for charged spheres extremality is only achieved when $M/R$ = $Q/R$ = 1. In our study we will look for cases where extremality can occur before the upper bound on $Q/R$ and $M/R$ is reached. 

This paper is organized as follows in the next section we develop the full Einstein-Maxwell equations for the interior gravitational field of a charged sphere. In section 3 we solve the equations and consider their stability properties and the formation of extremal black holes in these solutions.  In section 4 we discuss our results and formulate our conclusions. We note that a prime ($'$) or a comma ($,$) denotes derivatives with respect $r$, and a semi-colon ($;$) is used to represent covariant derivatives. We will work in units where $c = G = 1$.

The material presented in the next section  is well known. However, for consistency	 and clarity we have chosen to show the full derivation of the electromagnetic energy-momentum tensor for a static radial electric field.  A complete pedagogical description	of the derivation of the general form of the perfect fluid and electromagnetic energy-momentum tensors is given in Alder, Bazin and Schiffer \cite{Alder}.

\section{The field equations}
We are  interested in studying the interior gravitational field of  charged spheres, therefore we will assume a spherically symmetric metric of the form
\begin{equation}
ds^2  = - e^{2 \nu} dt^2 + e^{2 \lambda} dr^2 + r^2 d\theta^2 + r^2 \sin^2\theta d\phi^2.
\end{equation}
In this paper we are concerned only with  spherically symmetric static solutions  of the coupled Einstein-Maxwell equations, therefore $\nu \, = \, \nu(r)$ and $\lambda\,=\, \lambda(r)$ are functions of $r$ only.  
The Einstein field equations are 
\begin{equation}
\label{EF} 
G_{\alpha \beta} = R_{\alpha \beta} - \frac{1}{2} g_{\alpha \beta} R = 8 \pi T_{\alpha \beta}.
\end{equation}
 The spheres that we will study are composed of  a perfect fluid with a charge distribution that creates  a static radial electric field. The energy-momentum tensor $T_{\alpha \beta}$,  will thus be written as
\begin{equation}
T_{\alpha \beta}  = (T_{\alpha \beta})_{pf} + (T_{\alpha \beta})_{em},
\end{equation}
with $(T_{\alpha  \beta})_{pf}$  the energy-momentum tensor for a perfect fluid and  $  (T_{\alpha \beta})_{em}  $ the energy-momentum tensor   associated with the electric field. 

The energy momentum tensor  for a static spherically symmetric perfect fluid with a rest-frame energy-mass density $\rho$  and isotropic  pressure $p$  \cite{Alder} is
\begin{equation}
\label{Tfl}
(T_{\alpha \beta})_{f} = ( \rho + p) u_{\alpha} u_{\beta} + p g_{\alpha \beta}.
\end{equation} 
Here,  $\rho$ and $p$ are functions  of $r$ only and $u_{\alpha}$ is the 4-velocity of the fluid.

The four velocity $u_{\alpha}$ is defined such that
\begin{equation}
g_{\alpha \beta} u^{\alpha} u^{\beta} = -1.
\end{equation}
Since we are considering a fluid at rest, we will take 
\begin{equation}
u_{r} = u_{\theta} = u_{\phi} = 0,  ~~~ {\rm and}~~~u_{t} = - (-g^{tt})^{-\frac{1}{2}} = - e^{\nu(r)}
\end{equation}
thus, 
\begin{equation}
\label {eq1}
u_{\alpha} = -\delta_{\alpha}^{t} e^{\nu}
\end{equation}
and the energy momentum tensor for a perfect fluid is
\begin{equation}
(T_{\alpha \beta })_{pf} = {\rm{diag}} (-\rho e^{-2\nu} , p \,e^{-2\lambda} , p \,r^2  , p\, r^2 \sin^2 \theta ).
\end{equation} 
The energy-momentum tensor of the electric field, $(T_{\alpha \beta})_{em}$ is constructed from the electromagnetic field tensor $F_{\alpha \beta}$:  
\begin{equation}
(T_{\alpha \beta })_{em}  = \frac{1}{4 \pi} ( F_{\alpha}^{~ \tau} F_{\beta \tau}  - \frac{1}{4}  g_{\alpha \beta } F_{\tau \kappa} F^{\tau \kappa} ).
\end{equation}
We are interested in studying a  fluid whose charge distribution gives rise to static radial electric field therefore the only non-zero components of $F_{\alpha \beta}$ are  $F_{rt} \, =\, -F_{tr}  = \, E(r)$, thus
\begin{equation}
\label {eq2}
F_{\alpha \beta } = \delta_{\alpha}^r \delta_{\beta}^t E(r).
\end{equation}
Maxwell's equations in  curved space-time are:
\begin{equation}
F^{\alpha \beta }_{~~~; \beta } = 4 \pi j^{\alpha}, ~~{\rm and~~} F_{[\alpha \beta ;\tau} + F_{[\tau \alpha; \beta ]} + F_{[\beta \tau; \alpha]} = 0.
\end{equation}
Given the form of $F_{\alpha \beta}$   (\ref{eq2}), the second set of equations is identically zero, also since $F_{\alpha \beta} $ is anti-symmetric the first set of equations can be written as 
\begin{equation}
\label{eq3}
\frac{1}{\sqrt{-g} }(\sqrt{-g} F^{\alpha \beta})_{, \beta} = 4 \pi j^{\alpha},
\end{equation}
where $j^{\alpha}$ is the current 4-vector. If the charge density $\sigma$ is given then 
\begin{equation}
j^{\alpha} = \sigma u^{\alpha}.
\end{equation} 
Using the explicit forms of  $F_{\alpha \beta }$ and $u_{\alpha }$  in equation (\ref{eq3}) we find that
\begin{equation}
(r^2 e^{-(\nu + \lambda)} E(r))_{,r} = 4 \pi  r^2 \sigma e^{\lambda}.
\end{equation}
Integrating this equation gives  an expression for $E(r)$:
\begin{equation}
\label{Er}
E(r) = \frac{e^{\nu+ \lambda} q(r)  }{r^2}, 
\end{equation}
with
\begin{equation}
\label{charge}
q(r) = \int_0^r 4 \pi  r^2 \sigma e^{\lambda} dr.
\end{equation} 
We note that since $\sigma = j^t e^{\nu}$, $q(r) $ can be written as 

\begin{equation}
q(r) =  \int_0^{r} \int_0^{\pi} \int_{0}^{2 \pi}   j^t   e^{\nu+\lambda}  r^2 \sin \theta        d\theta d \phi dr = \int \limits_V  j^t  \sqrt{-g} \,dV.
\end{equation}
Thus $q(r) $ represents the total charge contained in a  sphere of radius $r$.
Using (\ref{Er}) we can write 
\begin{equation}
F_{\alpha \beta }  =  \delta_{\alpha}^r \delta_{\beta}^t \frac{e^{\nu + \lambda} q(r)  }{r^2},
\end{equation}
and the electromagnetic energy-momentum tensor takes the from 

\begin{equation}
(T_{\alpha \beta})_{em}  = \frac{q(r) ^2}{8 \pi r^4} {\rm diag} ( e^{-2 \nu}, - e^{2\lambda}, r^2, r^2 \sin^2 \theta).
\end{equation}

We can now write out complete set of equations  that describe  the interior gravitational field of static charged spheres.  They are:
\begin{equation}
\label{gt}
G^t_{~t} = e^{-2 \lambda}  \left( \frac{2 \lambda^{\prime}} {r}  - \frac{1}{r^2} \right)  +  \frac{1}{r^2}     = 8 \pi \rho + \frac{q^2}{r^4} ,
\end{equation}

\begin{equation}
\label{grr}
G^r_{~r} = e^{-2 \lambda}  \left( \frac{2 \nu^{\prime}} {r}  + \frac{1}{r^2} \right)  -  \frac{1}{r^2}     = 8 \pi p - \frac{q^2}{r^4} ,
\end{equation}
and
\begin{equation}
\label{gth}
 G^\theta_{~\theta} = G^\phi_{~\phi }= e^{-2 \lambda}  \left(   \nu^{\prime \prime} +  {\nu^{\prime}}^2  - \nu^{\prime}\lambda^{\prime} + \frac{\nu^{\prime}} {r}  - \frac{\lambda^{\prime}}{r} \right)   
    = 8 \pi p + \frac{q^2}{r^4} .
\end{equation}

\section{Solutions for the field equations}
We will now develop solutions for the field equations. We start by noting that (\ref{gt}),  can be written as

\begin{equation}
\frac{d}{dr} (r e^{-2 \lambda} ) = 1 - 8 \pi \rho  r^2  - \frac{q^2}{r^2}.  
\end{equation}
This equation can be immediately integrated to give 
\begin{equation}
\label{eli}
e^{-2 \lambda(r)} = 1- \frac{2m_{i}(r)}{r} - \frac{f(r)}{r},
\end{equation}
with
\begin{equation}
m_{i}(r) =  4 \pi \int_0^{r}  \rho (r^{\prime})  {r^{\prime}}^2 d r^{\prime}   ~~~{\rm and}~~~ f(r) = \int_0^r \frac{q(r^{\prime})^2}{{r^{\prime}}^2}dr^{\prime}.
\end{equation}
The quantity $m_i(r) $ is the inertial mass of the  fluid in a sphere of radius $r$. A related quantity is the total gravitational mass $m_g(r) $  of the fluid in a sphere of radius $r$.  Bekenstein \cite{Bekenstein}  in his study of charged spheres introduced $m_g(r)$ in an $ad \,hoc$ manner. Since the exterior metric of a charged sphere, the Reissner-Nordstr\"{o}m solution has
 \begin{equation}
 \label{elg2} 
e^{-2 \lambda(r)} = 1- \frac{2M}{r} + \frac{Q^2}{r^2}
\end{equation} 
where $M$ is the total gravitational mass of the sphere and $Q$ is the total charge of the sphere, 
Bekenstein proposed that in the interior  of a charged sphere,  $e^{-2\lambda(r)}$ should have following form:
 \begin{equation}
 \label{elg} 
e^{-2 \lambda(r)} = 1- \frac{2m_{g}(r)}{r} + \frac{q^2(r)}{r^2}.
\end{equation} 

The function $q(r)$ is the total charge in a sphere of radius $r$. It is the  charge function defined in (\ref{charge})     that we have been studying. In  order for  (\ref{eli}) to be equal to (\ref{elg}), $m_g(r)$ must be defined in the following manner:
\begin{equation}
m_g(r) \equiv \frac{1}{2 } \int_0 ^r \left(8 \pi \rho {r^{\prime}}^2 + \frac{q^2(r^{\prime})}{{r^{\prime}}^2} \right)  d r^{\prime} + \frac{q(r)^2}{2 r}.
\end{equation}
We note that in the absence of the electric field $m_g = m_i$.

The requirement  that $e^{- 2 \lambda(r) }$ matches  the Reissner-Nordstr\"{o}m metric  at surface of the sphere, $r = R$ gives 
\begin{equation}
\label{mass}
1 - \frac{M}{R} + \frac{Q^2}{R^2}  = 1  - \frac{1}{R} \int_0^R ( 8 \pi \rho r^2  + \frac{q^2}{r^2} ) dr.
\end{equation}
An expression for the total mass can be found from this equation:
\begin{equation}
M =  \frac{1}{2} \int_0^R ( 8 \pi \rho r^2  + \frac{q^2}{r^2} ) dr  +  \frac{Q^2}{2R}.
\end{equation}
In this paper we will study charged spheres with the following inertial mass density $\rho(r)$ and charge distribution $q(r)$ profiles:
\begin{equation}
\label{rhoa}
\rho(r) = \rho_o - \frac{b  }{8 \pi}  \frac{Q^2}{R^6} r^2  {\rm ~~~~and~~~~}  q(r) =   \frac{Q}{R^3} r^3,
\end{equation}
in the expression for $\rho(r)$, $b$ is a number. In our models 
\begin{equation}
\label{elambda}
e^{-2 \lambda(r)} = 1- \frac{8 \pi \rho_o}{3} r^2  + \frac{( b  -1)}{5}  \frac{Q^2}{R^6} {r^4}
\end{equation}
\begin{equation}
\label{mgeq}
m_g(r)  =  \frac{4 \pi \rho_o}{3}r^3 +   \frac{(6 - b )}{10}\frac{Q^2}{2 R} \frac{r^5}{R^5}.
\end{equation}
and
\begin{equation}
\label{Meq}
M =  \frac{4 \pi \rho_o}{3}R^3 +   \frac{(6 - b )}{10}\frac{Q^2}{2 R}.
\end{equation}
With the assumed mass  and charge distributions  here,  when $b = 6$ 
\begin{equation}
 \frac{m_g(r)}{r^3} = \frac{M}{R^3} = const,
\end{equation}
thus the $b = 6$ model is a sphere with a constant gravitational mass density. Also all  models have 
\begin{equation}
\frac{ q(r) }{r^3} =  \left( \frac{Q}{R^3}\right)  = const.
 \end{equation}
 thus the charge density is constant for all our models. 
The models that we will study in detail here include following three charged configurations:
\begin{enumerate}
\item  $b = 0$  - a sphere with a constant  inertial mass density  and constant charge density. 
\item  $b = 1$ -  a sphere with constant total energy density and constant charge density.
\item $b = 6 $ - a sphere with constant gravitational mass density and constant charge density. 
\end{enumerate}
We can solve for $\rho_o$ from (\ref{Meq}) and rewrite (\ref{elambda}) as
\begin{equation}
\label{lambda}
e^{-2 \lambda(r)} = 1-  \left( \frac{2M}{R}  + \frac{(b-6)}{5}  \frac{Q^2}{R^2}    \right) \frac{r^2}{R^2}  + \frac{(b-1)}{5}  \frac{Q^2}{R^2} \frac{r^4}{R^4}.
\end{equation}
Thus, if we  are given $\rho(r) $ and $q(r)$ we have found $ \lambda(r)$. 

We now need to solve for $\nu(r)$. We start by transforming (\ref{gth}). First we subtract   (\ref{grr})  from  (\ref{gth} )  to get
\begin{equation}
 e^{-2 \lambda}  \left(   \nu^{\prime \prime} +  {\nu^{\prime}}^2  - \nu^{\prime}\lambda^{\prime} - \frac{\nu^{\prime}} {r}   \right)  
 -  \frac{\lambda^{\prime} e^{-2 \lambda}}{r}                         -  \frac{ e^{-2 \lambda}}{r^2}  +  \frac{1}{r^2}                                                                              = 2 \frac{q^2}{r^4} .
\end{equation}
Then we substitute for $1/r^2$ from   (\ref{gt})   to get  the following equation
\begin{equation}
\label{w1}
  \nu^{\prime \prime} +  {\nu^{\prime}}^2  - \nu^{\prime}\lambda^{\prime} - \frac{\nu^{\prime}} {r}   
= 3 \frac{ \lambda^{\prime}} {r}  -  \left( 8 \pi \rho  - \frac{q^2}{r^4} \right)   e^{2\lambda} .
 \end{equation}
We now multiply both sides of this equation  by $e^{- \lambda + \nu}/r$, and we find that the left hand-side becomes an exact differential: $ ((\nu^{\prime} e^{- \lambda + \nu})/r)^{\prime}$. Introducing $\zeta(r)\equiv e^{\nu(r)}$, we can write (\ref{w1}) in the following form 
\begin{equation}
\label{w2}
\left( \frac{1}{r} e^{-\lambda} \zeta^{\prime} \right)^{\prime}  = \left[ \frac{3 \lambda^{\prime} e^{-2 \lambda}} {r^2}  -   \frac{8 \pi \rho}{r}  + \frac{q^2}{r^5}   \right] e^{\lambda} \zeta.
\end{equation} 

This equation was derived by Giuliani and Rothman \cite {Giuliani} in their study of charged spheres.  The transformation of the left hand-side  (\ref{w1}) into  that of (\ref{w2}) was introduced by Weinberg \cite{Weinberg} in deriving the equation he used to prove the Buchdahl limit of $M/R$ for neutral perfect fluid spheres.  

Substituting for $\rho$, $q$ (from (\ref{rhoa})) and $\lambda^{\prime}e^{- 2 \lambda}$ (from (\ref{lambda}))  we find that (\ref{w2}) becomes 
\begin{equation}
\label{zeta}
\left( \frac{1}{r} e^{-\lambda} \zeta^{\prime} \right)^{\prime}  = \left(\frac{11}{5} - \frac{b}{5}\right) \frac{Q^2}{R^6} r e^{\lambda} \zeta.
\end{equation}
We now define a new variable 
\begin{equation}
\label{utransform}
\tilde{\zeta}( u(r))  \equiv \zeta(r)  {\rm ~~~~~with~~~~}  u(r)  = \frac{1}{R^2} \int_0^r s e^{\lambda(s) } ds.
\end{equation}
Then (\ref{zeta}) becomes
\begin{equation}
\label{Th}
\frac{ d^2 \tilde{\zeta}}{d u^2}  =     \left(\frac{11}{5} - \frac{b}{5}\right)          \frac{Q^2}{R^2}  \tilde{\zeta} (u) . 
\end{equation}
This is our master equation. Our task now is to find solutions for it given various values of $b$. There are three types of solutions of the master equation depending on the value of $b$:
 \begin{equation} 
 \label{k1eq}
 \fl (i)~~  b <  11:  ~~ \tilde{\zeta} (u(r)) = A e^{a \frac{Q}{R} u(r)}  +  B  e^{-a \frac{Q}{R} u(r)}   ~~~{\rm with}~ a =  {\left(\frac{11}{5} - \frac{b}{5}\right)  }         ^{\frac{1}{2}},
\end{equation}
\begin{equation}
\label{k2eq} 
\fl (ii) ~~b = 11:  ~~\tilde{\zeta} (u(r))   = A  +  B  u(r) ,
\end{equation}
\begin{equation}
\label{k3eq}
\fl  (iii)~~ b >   11 : ~~  \tilde{\zeta} (u(r)) = A \sin\left(d \frac{Q}{R} u(r)\right)  +  B  \cos \left(-d \frac{Q}{R} u(r)\right)  ~~~{\rm with} ~d ={\left|\frac{11}{5} - \frac{b}{5}\right|}^{\frac{1}{2}}.
 \end{equation}

\noindent The values of the constants $A$ and $B$ in (\ref{k1eq}), (\ref{k2eq})  and (\ref{k3eq}) are fixed by  the boundary conditions imposed on $\tilde{\zeta}(u)$:
\begin{equation}
\tilde{\zeta}(u(R)) =  \left(1 - \frac{2M}{R} + \frac{Q^2}{R^2}\right)^{\frac{1}{2}} 
\end{equation}
and
\begin{equation}
\frac{d\tilde{\zeta}(u(R))}{du(R) }  =  \left(\frac{M}{R} - \frac{Q^2}{R^2}\right).
\end{equation}
The condition for stability is $\zeta(r = 0) > 0$. The critical values of $M/R$ as a function of $Q/R$ are found from solving the equation $\zeta(r=0) = 0$.

\subsection {The $ b =  0 $ solution.}
We start developing solutions for  (\ref{Th}) by studying the case with  $b =  0$. This choice of $b$ corresponds to a fluid with constant inertial density and a constant charge density. This model will serve as an important prototype for comparing the  effects of changing $b$ on the physical properties of the system. 
When $b = 0$,  we find that 

\begin{equation}
e^{-2 \lambda(r)} = 1-  \left( \frac{2M}{R}  - \frac{6}{5}  \frac{Q^2}{R^2}    \right) \frac{r^2}{R^2}  -  \frac{1}{5}  \frac{Q^2}{R^2} \frac{r^4}{R^4},
\end{equation}

\begin{eqnarray}
\fl~~~~~~~~~ u(r) =    \frac{1}{R^2} \int_0^r   e^{\lambda(s)}  s ds                               = \frac{1}{R^2} \int_0^r \frac{s ds }{  \left[1-  \left( \frac{2M}{R}  - \frac{6}{5}   \frac{Q^2}{R^2}    \right) \frac{s^2}{R^2}    -\frac{1}{5} \frac{Q^2}{R^2} \frac{s^4}{R^4} \right]^{\frac{1}{2}}}\\
~~~~~~~~~~~~~~~\fl = \frac{\sqrt{5}}{2}\frac{R}{Q} \left[ \tan^{-1}  \left(  \left(     \sqrt{5} \frac{M}{Q} -  \frac{3}{\sqrt{5}}\frac{Q}{ R }+  \frac{1}{\sqrt{5}}  \frac{Q }{R^3} r^2 \right) e^{\lambda(r)} \right)  \right. \\ \nonumber
~~~~~~~~~~~~~~~~~~~~~~~~~~~~~~~~~~~~~~~~~~~~~~\left.  -\tan^{-1}       \left(  \sqrt{5} \frac{M}{Q} -  \frac{3}{\sqrt{5}}\frac{Q}{ R }\right)\right]
\end{eqnarray}
and the master equation becomes
\begin{equation}
\label{k0}
\frac{ d^2 \tilde{\zeta}}{d u^2}  = \frac{11}{5}  \frac{Q^2}{R^2}  \tilde{\zeta} (u).
\end{equation}
The solution for this equation is  
\begin{equation}
\tilde{\zeta} (u(r))  = A  \exp \left( \sqrt{\frac{11}{5}} ~\frac{Q}{R} u(r)   \right)          + B \exp \left(-{  \sqrt{\frac{11}{5}}   ~   \frac{Q}{R} u(r)  } \right).
 \end{equation}
The constants $A$ and $B$ are found using the boundary conditions. Solving  for them we find we find, 
\begin{equation}
\fl ~~~~~~~~~A = \frac{1}{2} \left[ \left(1 - \frac{2M}{R} + \frac{Q^2}{R^2}\right)^{\frac{1}{2}}  +    \sqrt{\frac{5}{11}}   \left(\frac{M}{Q} - \frac{Q}{R}\right)  \right]  \exp \left(-{  \sqrt{\frac{11}{5}}     ~     \frac{Q}{R} u(R)  } \right)
\end{equation}
and
\begin{equation}
\fl ~~~~~~~~B = \frac{1}{2} \left[ \left(1 - \frac{2M}{R} + \frac{Q^2}{R^2}\right)^{\frac{1}{2}}  -  \sqrt{\frac{5}{11}}      \left(\frac{M}{Q} - \frac{Q}{R}\right)  \right]  \exp \left({ \sqrt{\frac{11}{5}}   ~      \frac{Q}{R} u(R)  } \right).
\end{equation}

\begin{figure}
\centering
 \includegraphics[width=0.75\linewidth]{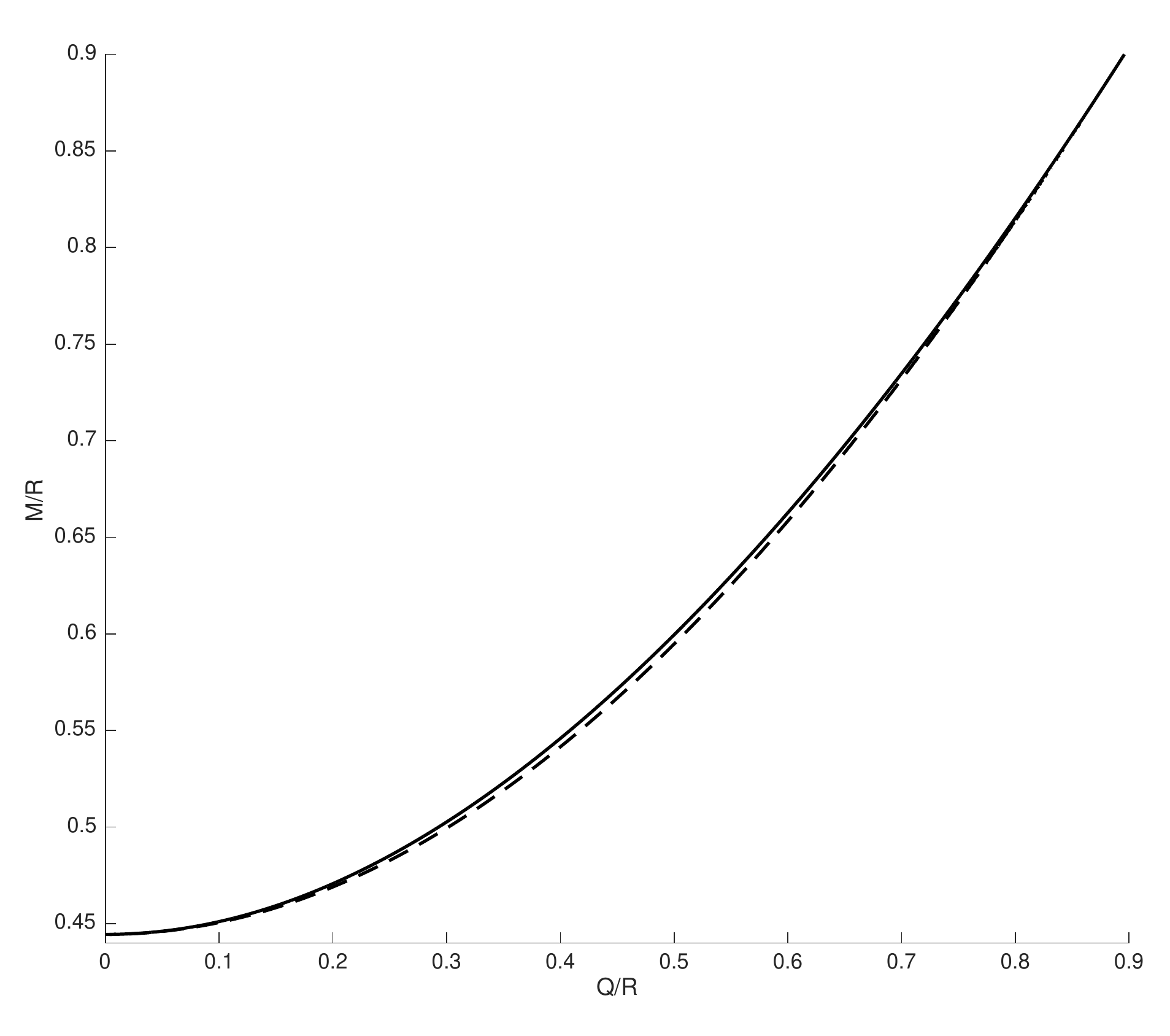}
 \caption{ \it  The critical values of M/R vs Q/R  for   $b =  0$   $({\bf ---})$   and     the  Andr\'{e}asson formula   (\rule{0.9 cm}{0.03 cm}).      }
\label{fig:f1}
\end{figure}

The critical values of $M/R$ vs $Q/R$ are found from condition $\tilde{\zeta}(u(r=0)) = 0$. Here this condition  requires $A + B = 0$ and leads to the following transcendental equation for $M/R$ as function of $Q/R$:
\begin{eqnarray}
\fl\frac {\left[     \sqrt{\frac{5}{11}} \left(\frac{M}{Q} - \frac{Q}{R}\right)           +      \left(1 - \frac{2M}{R} + \frac{Q^2}{R^2}\right)^{\frac{1}{2}}               \right]} 
{ \left[        \sqrt{\frac{5}{11}}         \left(\frac{M}{Q} - \frac{Q}{R}\right)           -      \left(1 - \frac{2M}{R} + \frac{Q^2}{R^2}\right)^{\frac{1}{2}}           \right] }  = \\ \nonumber
\fl ~~~~~~~~~~~~ \exp \left (\sqrt{11} \left( \left[ \tan^{-1}  \left(  \left(     \sqrt{5} \frac{M}{Q} -  \frac{2}{\sqrt{5}}\frac{Q}{ R } \right) e^{\lambda(R)} \right) \right. \right. \right. 
\left. \left. \left.  -\tan^{-1}       \left(  \sqrt{5} \frac{M}{Q} -  \frac{3}{\sqrt{5}}\frac{Q}{ R }\right)\right] \right)\right)
\end{eqnarray}
 We solved this equation numerically  and the results are plotted in Figure \ref{fig:f1}. For comparison we also plotted the corresponding values of $M/R$ vs $Q/R$ from Andr\'{e}asson's formula. We find that   critical values of $M/R$  from this model almost saturates the Andr\'{e}asson limit but does not exceed it.   We note that we were able to integrate only up to $Q/R \approx 0.9$ with this model.  
 
 \subsection{ The $ b = 1$ solution.}
 When $b = 1$,
\begin{equation}
e^{-2 \lambda(r)} = 1-  \left( \frac{2M}{R}  -   \frac{Q^2}{R^2}    \right) \frac{r^2}{R^2},
\end{equation}
\begin{eqnarray}
\fl~~~~~~~~~~~~u(r) =  \frac{1}{R^2} \int_0^r   e^{\lambda(s)}  s ds  =  \left( \frac{2M}{R} - \frac{Q^2}{R^2} \right)^{-1}  \left[ 1 -  \left[  1-  \left( \frac{2M}{R}  -   \frac{Q^2}{R^2}    \right) \frac{r^2}{R^2}\right]^{\frac{1}{2}} \right].
\end{eqnarray}
and the master equation becomes 
 \begin{equation}
\label{k1}
\frac{ d^2 \tilde{\zeta}}{d u^2}  = 2 \frac{Q^2}{R^2}  \tilde{\zeta} (u).
\end{equation} 
The solutions here are 
\begin{equation}
\tilde{\zeta} (u(r))  = A  \exp\left({\sqrt{ 2} \frac{Q}{R} u(r)  } \right)          +  B \left({-\sqrt{ 2} \frac{Q}{R} u(r)  }\right),
 \end{equation}
 with
 \begin{equation}
A = \frac{1}{2} \left[ \left(1 - \frac{2M}{R} + \frac{Q^2}{R^2}\right)^{\frac{1}{2}}  +  {\frac{1}{\sqrt{2}}}      \left(\frac{M}{Q} - \frac{Q}{R}\right)  \right] \exp\left({-\sqrt{2} \frac{Q}{R} u(R)  }\right),
\end{equation}

\begin{equation}
B = \frac{1}{2} \left[ \left(1 - \frac{2M}{R} + \frac{Q^2}{R^2} \right)^{\frac{1}{2}}  -    {\frac{1}{\sqrt{2}}}     \left(\frac{M}{Q} - \frac{Q}{R}\right)  \right] \exp\left({\sqrt{ 2} \frac{Q}{R} u(R)  }\right)
\end{equation}
and
\begin{eqnarray}
u(R) 
=  \left( \frac{2M}{R} - \frac{Q^2}{R^2} \right)^{-1}  \left[ 1 -  \left(  1-  \frac{2M}{R}  +   \frac{Q^2}{R^2}    \right)^{\frac{1}{2}} \right].
\end{eqnarray}
Here the critical values of $M/R$ are found from the following equation:
\begin{eqnarray}
\label{aneq33}
\frac{\left[        \frac{1}{\sqrt{2}}     \left(\frac{M}{Q} - \frac{Q}{R}\right)           +      \left(1 - \frac{2M}{R} + \frac{Q^2}{R^2}\right)^{\frac{1}{2}}               \right]}{ \left[        \frac{1}{\sqrt{2}}     \left(\frac{M}{Q} - \frac{Q}{R}\right)           -      \left(1 - \frac{2M}{R} + \frac{Q^2}{R^2}\right)^{\frac{1}{2}}           \right] }  =  \\ \nonumber
~~~~~~~~~~~~~~~~~~~~~~~~~~~~~~~~~~~\exp \left [ \frac{2 \sqrt{2} }{\left( \frac{2M}{Q} - \frac{Q}{R} \right)} 
 \left[ 1 -  \left(  1-  \frac{2M}{R}  +   \frac{Q^2}{R^2}    \right)^{\frac{1}{2}} \right] \right]
\end{eqnarray}

 \begin{figure}
\centering
 \includegraphics[width=0.75\linewidth]{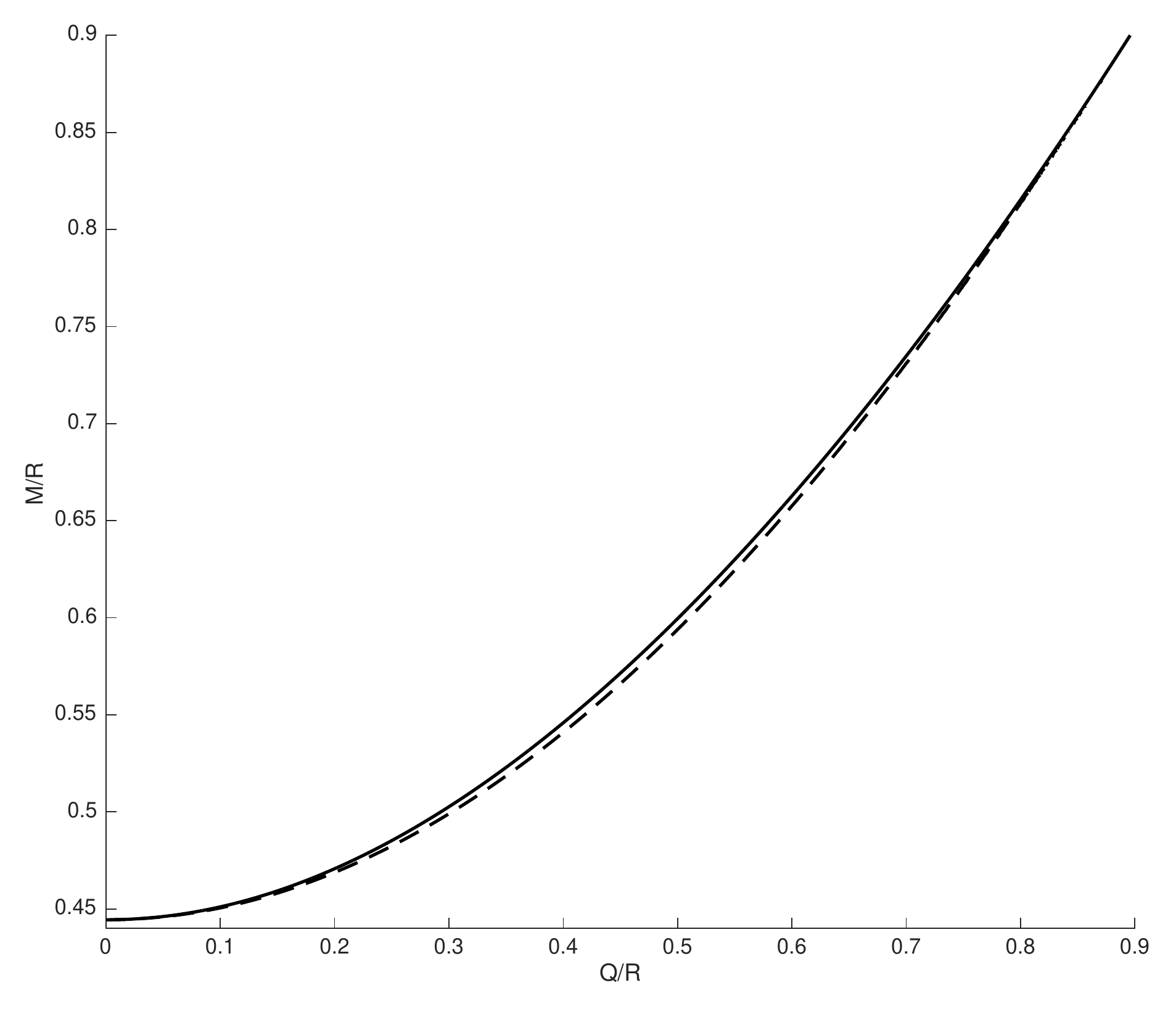}
 \caption{\it  The critical values of M/R vs Q/R  for   $b =  1$   $({\bf ---})$   and     the  Andr\'{e}asson formula   (\rule{0.9 cm}{0.03 cm}).      }
\label{fig:f2}
\end{figure} 

The critical values from (\ref{aneq33}) for the $b = 1$ model and the  Andr\'{e}asson formula are compared in figure \ref{fig:f2}.
 We see that   $ b = 1$ the critical values are less than the corresponding values from the Andr\'{e}asson  formula.   

 \subsection {The $ b = 6$  solution.}
 When $b = 6$,
 \begin{equation}
e^{-2 \lambda(r)} = 1-  \frac{2M}{R}  \frac{r^2}{R^2}  +  \frac{Q^2}{R^2} \frac{r^4}{R^4},
\end{equation} 
 \begin{eqnarray}
 \fl ~~~ u(r) =  \frac{R}{2 Q} \left [ \log \left(\frac{M}{Q} + 1\right)  - \log \left( \frac{M}{Q} -   \frac{Q}{R^3} r^2 +  \left(  1-  \frac{2M}{R}  \frac{r^2}{R^2}  +  \frac{Q^2}{R^2} \frac{r^4}{R^4 } \right)^{\frac{1}{2}}                  \right) \right]
\end{eqnarray} 

and the master equation becomes 
 \begin{equation}
\label{8pik6}
\frac{ d^2 \tilde{\zeta}}{d u^2}  =  \frac{Q^2}{R^2}  \tilde{\zeta} (u),
\end{equation}

 The solution for (\ref{8pik6}) is 
 \begin{equation}
\tilde{\zeta} (u(r))  = A  e^{ \frac{Q}{R} u(r)  }           +  B e^{- \frac{Q}{R} u(r)  }.
 \end{equation} 
 The boundary conditions require that 
 \begin{equation}
 \left(1 - \frac{2M}{R} + \frac{Q^2}{R^2}\right)^{\frac{1}{2}} = A  e^{ \frac{Q}{R} u(R)  }           +  B e^{- \frac{Q}{R} u(R)  }.
 \end{equation}
and
\begin{equation}
  \left(\frac{M}{R} - \frac{Q^2}{R^2}\right) =   \frac{Q}{R} \left(  A  e^{ \frac{Q}{R} u(R)  }           -  B e^{- \frac{Q}{R} u(R)  } \right).
\end{equation} 

 Solving for $A$ and $B$  we find that 
 \begin{equation}
 A =  \frac{1}{2}   \left(\frac{M}{Q} + 1\right)^{-\frac{1}{2}}        \left[\frac{M}{Q} - \frac{Q}{R} + \left(1 - \frac{2M}{R} + \frac{Q^2}{R^2}\right)^{\frac{1}{2}} \right]^{\frac{3}{2}}  
 \end{equation}
 and
  \begin{eqnarray}
  B = \frac{1}{2}     \left(\frac{M}{Q} + 1\right)^{\frac{1}{2}}   
    \frac{\left[\frac{Q}{R} - \frac{M}{Q} +  \left(1 - \frac{2M}{R} + \frac{Q^2}{R^2}\right)^{\frac{1}{2}}   \right]}
  {\left[\frac{M}{Q} - \frac{Q}{R} +  \left(1 - \frac{2M}{R} + \frac{Q^2}{R^2}\right)^{\frac{1}{2}}   \right]^{\frac{1}{2}}}.
 \end{eqnarray} 
 The stability condition $\zeta( r = 0) = 0$ leads to the following critical values equation:
 \begin{eqnarray}
  \fl      \left[\frac{M}{Q} - \frac{Q}{R} +   \left(1 - \frac{2M}{R} + \frac{Q^2}{R^2}\right)^{\frac{1}{2}}          \right]^{2} =     \left(\frac{M}{Q} + 1\right)    \left[\frac{M}{Q} - \frac{Q}{R} - \left(1 - \frac{2M}{R} + \frac{Q^2}{R^2}\right)^{\frac{1}{2}} \right]  \\
\fl ~~~~~~~~~~{\rm or~~~} 2\left(4 \frac{Q^4}{R^4} - 12\frac{M}{R}\frac{Q^2}{R^2} + 9\frac{M^2}{R^2} + 3 \frac{Q^2}{R^2}  - 4\frac{M}{R}\right)\frac{Q^2}{R^2} \left(\frac{M}{R} + \frac{Q}{R}\right) = 0.
 \end{eqnarray} 
 The term that is quadratic in $M/R$ has exact solutions for  $M/R$  as a function of $Q/R$. These  solutions are 
 \begin{equation}
\fl~~~~~~~ \frac{M}{R}  =  \frac{2}{9} +  \frac{2}{3} \frac{Q^2}{R^2}  - \frac{1}{9}\left(4 - \frac{3Q^2}{R^2}\right)^{\frac{1}{2}} {~~\rm and~~~} \frac{M}{R}  = \frac{2}{9} +  \frac{2}{3} \frac{Q^2}{R^2}  + \frac{1}{9}\left(4 - \frac{3Q^2}{R^2}\right)^{\frac{1}{2}}.
 \end{equation}
 The first of these two solutions gives $M/R = 0$ when $ Q/R = 0$. This result does not correspond to properties the model that we are studying here. Here, when $ Q/R = 0$,
  $M/R$ should be equal to $ 4/9$, the Buchdahl limit for neutral  perfect fluids. The second solution does give the correct  $M/R  = 4/9$ value when $ Q/R = 0$. Thus we will take it to represent the critical values for the $b = 6$ model.

 This model was studied extensively by Giuliani and Rothman \cite{Giuliani}. However they arrived at this model using a different approach from us. It is useful here to review their development of this model.  Giuliani and Rothman \cite{Giuliani} showed that \ref{w2} can be written in the following form:
 \begin{equation}
 \label{gandr}
\left( \frac{1}{r} e^{-\lambda} \zeta^{\prime} \right)^{\prime}  = \left[ \left (\frac{m_g(r)}{r^3} \right)^{\prime}   - q(r) \left(\frac{q(r)}{r^4}\right)^{\prime} \right] e^{\lambda} \zeta.
\end{equation}
They then proposed the following ansatz: $m_g = M (r/R)^3$ and $q = Q(r/R)^3$. The substitution of  their ansatz into (\ref{gandr}) brings into the following form:
 \begin{equation}
\left( \frac{1}{r} e^{-\lambda} \zeta^{\prime} \right)^{\prime}  = r \frac{Q^2}{R^6} e^{\lambda} \zeta.
\end{equation}
This is our (\ref{zeta}) for $b = 6$. 

 \begin{figure}
\centering
 \includegraphics[width=0.75\linewidth]{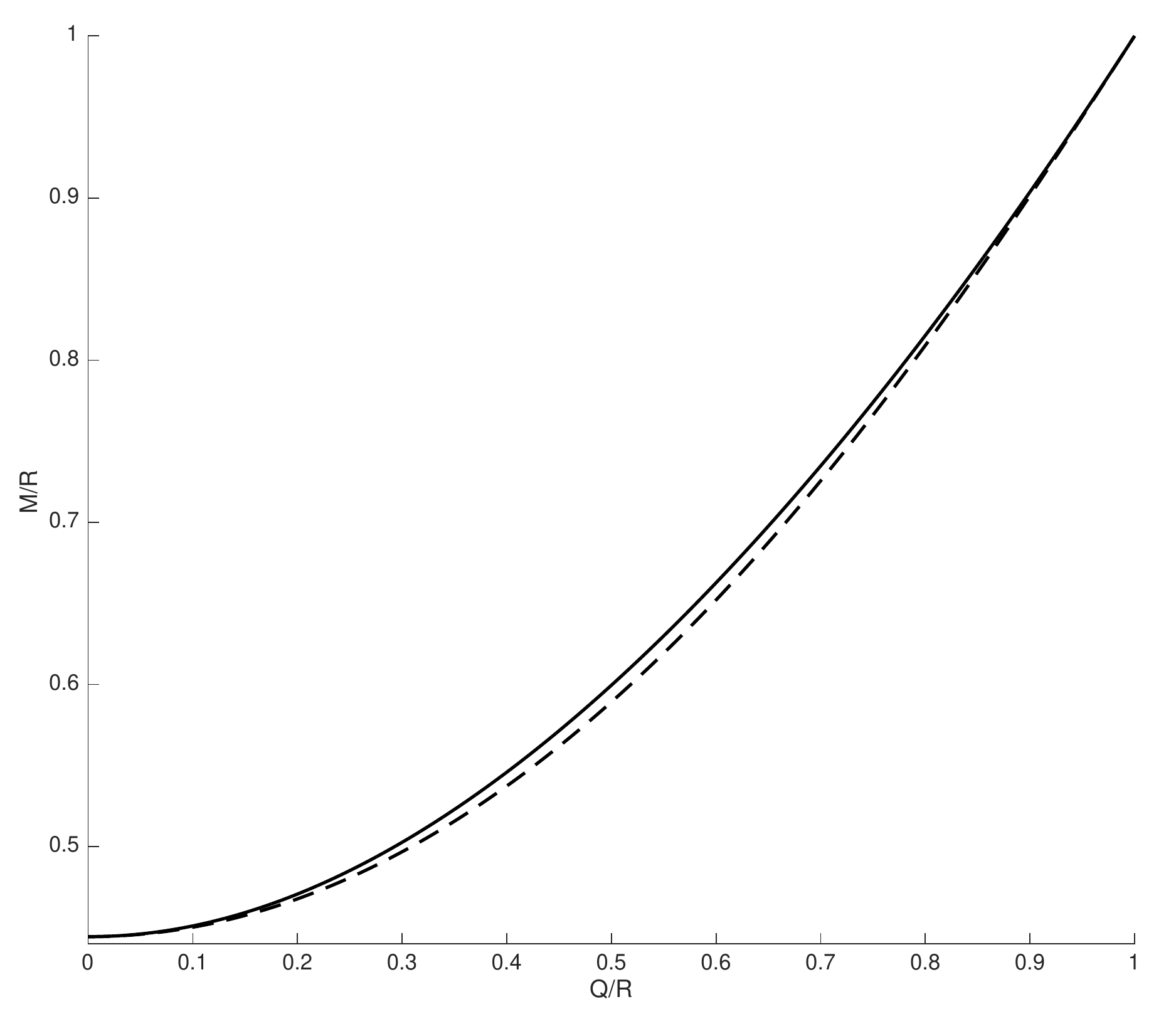}
 \caption{ \it The critical values of M/R vs Q/R  for   $b =  6$   $({\bf ---})$   and     the  Andr\'{e}asson formula   (\rule{0.9 cm}{0.03 cm}).      }
\label{fig:f3}
\end{figure} 

 \begin{figure}
\centering
 \includegraphics[width=0.75\linewidth]{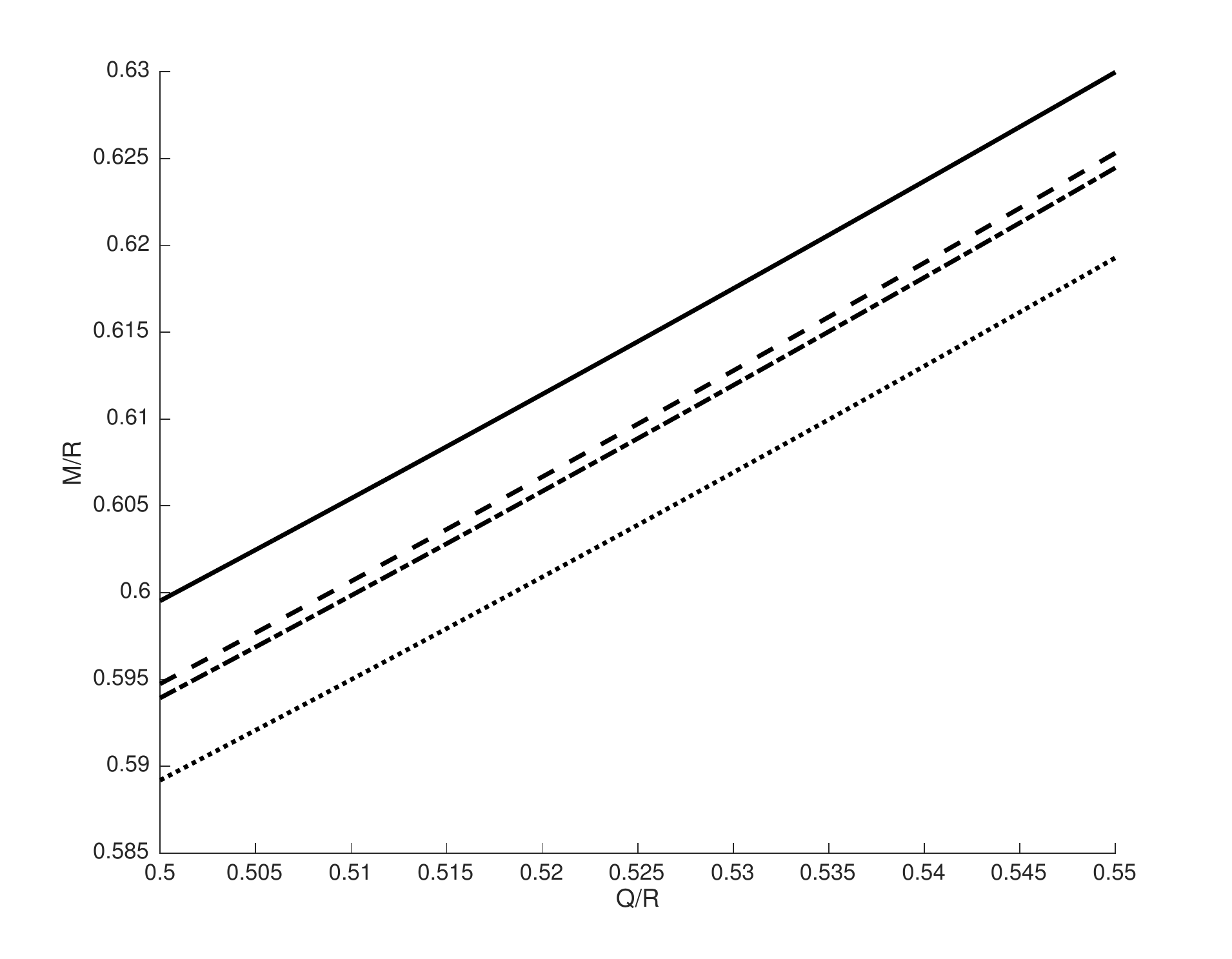}
 \caption{ \it The critical values of M/R vs Q/R  for  the  $b =  0$   $({\bf ----})$,  the $b =  1$   $({\bf - \cdot- \cdot-\cdot})$,   and the   $b =  0$   $({\bf \cdot\cdot\cdot\cdot\cdot})$ models                    and     the  Andr\'{e}asson formula   (\rule{0.9 cm}{0.03 cm}).      }
\label{fig:grb15}
\end{figure}

We  plotted the critical values of   $M/R$ vs $Q/R$  for this model in Figure \ref{fig:f3} along with the values from  the  Andr\'{e}asson formula. We find that this model ($b = 6$) does not saturate the  Andr\'{e}asson bound. At first glance, the critical values curves for the $b = 0$, the $b = 1$ and the $b = 6$  models do not appear differ from each other. In order to clearly show that the critical values of $M/R$ from these models are different from each other we plotted the results for all three models in figure \ref{fig:grb15} with the axes magnified. We find from figure \ref{fig:grb15}  that  for  spheres with a constant charge density, a  sphere with constant  inertial mass density (the $b = 0$ model) is more stable than a sphere with constant total energy density (the $b = 1$ model) which is turn more stable than  a  sphere with constant gravitational mass density (the $b = 6)$ model.

\subsection{The $b = 11$ solution}
For $b = 11$ 
 \begin{equation}
e^{-2 \lambda(r)} = 1-  \left (\frac{2M}{R}  +  \frac{Q^2}{R^2}\right )   \frac{r^2}{R^2} +  2 \frac{Q^2}{R^2}   \frac{r^4}{R^4}
\end{equation}
\begin{eqnarray}
\fl  u(r) = \frac{\sqrt{2}}{4}    \frac{R}{Q} \left[    \ln\left[ \sqrt{2} \left( \frac{M}{2 Q} + \frac{Q}{4R}  \right)
 +    1                          \right]  -  \ln\left[ \sqrt{2} \left( \frac{M}{2 Q} + \frac{Q}{4R} - \frac{Qr^2}{R^3} \right)
 +     e^{-\lambda(r)}                                   \right]  \right]
\end{eqnarray}
and the master equation is
\begin{equation}
\label{k11}
\frac{ d^2 \tilde{\zeta}}{d u^2}  = 0.
\end{equation}
The solution here is 
\begin{equation}
\tilde{\zeta}(u) = A + B u(r)
\end{equation}

\noindent   Applying the boundary conditions we find that 
 
 \begin{eqnarray}
 \fl \zeta(r) =  \frac{\sqrt{2}}{4}  \left( \frac{M}{Q} -   \frac{Q}{R}  \right)     \ln\left[ \frac{\sqrt{2} \left( \frac{M}{2 Q} - \frac{3Q}{4R}  \right) + e^{-\lambda(R)} } {  \sqrt{2} \left( \frac{M}{2 Q} + \frac{Q}{4R} - \frac{Qr^2}{R^3} \right)+     e^{-\lambda(r)} }\right]+             \left(1 - \frac{2M}{R} + \frac{Q^2}{R^2}\right)^{\frac{1}{2}}.
\end{eqnarray}
The critical values equation here is
\begin{eqnarray}
  \sqrt{ 1 - \frac{2M}{R} + \frac{Q^2}{R^2}}  =  \frac{\sqrt{2}}{4}  \left(  \frac{Q}{R}        - \frac{M}{Q}       \right)   \ln\left[\frac{ \sqrt{2} \left( \frac{M}{2 Q} - \frac{3Q}{4R}  \right)
 + e^{-\lambda(R)} } {\sqrt{2}  \left( \frac{M}{2 Q} + \frac{Q}{4R}  \right)
 + 1} \right].  
 \end{eqnarray}

 \begin{figure}
\centering
 \includegraphics[width=0.75\linewidth]{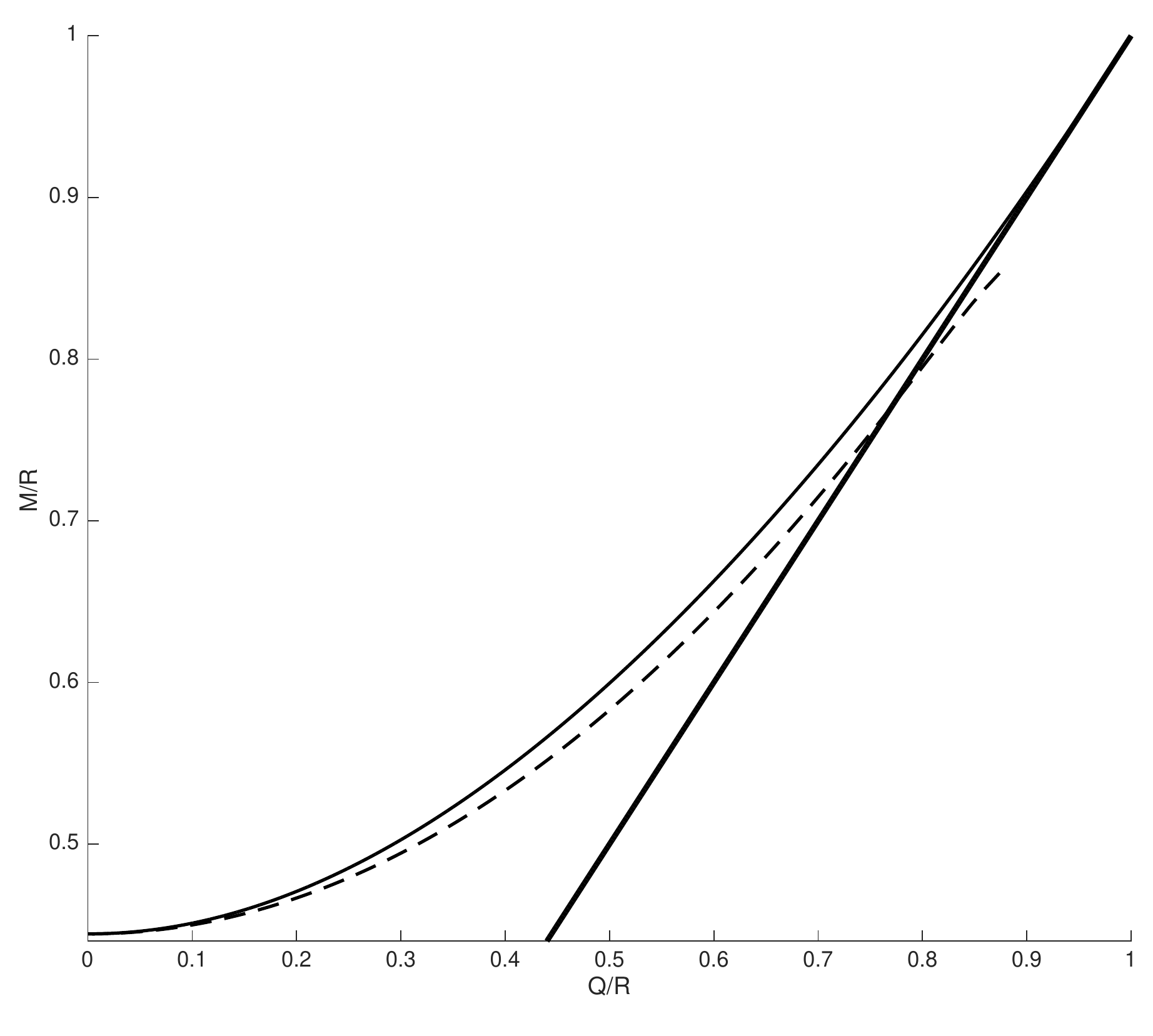}
 \caption{ \it  The critical values of M/R vs Q/R  for   $b =  11$   $({\bf ---})$   and     the  Andr\'{e}asson formula   (\rule{0.9 cm}{0.03 cm}).  The extremal line $M/R$ = $Q/R$ is also drawn (\rule{0.9 cm}{0.06 cm}).}
\label{fig:f8}
\end{figure} 

A plot the critical values of $M/R$ vs $Q/R$ for this model is given in Figure \ref{fig:f8}.  We find from Figure \ref{fig:f8}   that the critical values of $M/R$ in this model are  less than the values given by the   Andr\'{e}asson   limit. We were only able to integrate up to $Q/R \approx 0.88$. We also  found here that the extremal value $M/R = Q/R$ is achieved before the $Q/R = 1$ limit. Here  $M/R = Q/R$ when $Q/R = 0.77$.

\subsection{The $b = 16$ solution}
 When $b \geq 11 $,  the master equation becomes 
 \begin{equation}
\label{8pik16}
\frac{ d^2 \tilde{\zeta}}{d u^2}  =    - d^2 \frac{Q^2}{R^2}  \tilde{\zeta} (u)
\end{equation} 
with $d  = {|(11 - b)/5|}^{\frac{1}{2}}$ and the solution  is
 \begin{equation}
\tilde{\zeta} (u(r))  = A  \sin\left(d\frac{Q}{R} u(r) \right )           +  B \cos \left( d\frac{Q}{R} u(r)  \right),
 \end{equation} 
 with 
 \begin{equation}
 A =     e^{-\lambda(R)}        \sin\left(d\frac{Q}{R} u(R) \right )         +  \frac{1}{d} \left(\frac{M}{Q} - \frac{Q}{R} \right) \cos\left(d\frac{Q}{R} u(R) \right )  ,
\end{equation}

\begin{equation}
  B =     e^{-\lambda(R)}     \cos\left(d\frac{Q}{R} u(R) \right )        - \frac{1}{d} \left(\frac{M}{Q} - \frac{Q}{R} \right) \sin\left(d\frac{Q}{R} u(R) \right ).
\end{equation}

\noindent The stability condition $\zeta(r= 0) = 0$ requires
\begin{equation}
 A  \sin\left(d\frac{Q}{R} u(0) \right )           +  B \cos \left( d\frac{Q}{R} u(0)  \right) = 0,
 \end{equation} 
however, since $u(0) \equiv 0$, then  $B$ must be equal to zero here. This results in   the following critical values equation:
\begin{equation}
\label{366}
 e^{-\lambda(R)}     \cos\left(d\frac{Q}{R} u(R) \right )        = \frac{1}{d} \left(\frac{M}{Q} - \frac{Q}{R} \right) \sin\left(d\frac{Q}{R} u(R) \right ) 
 \end{equation}
Here for $b = 16$
\begin{equation}
e^{-2 \lambda(r)} = 1-  \left (\frac{2M}{R}  +   \frac{2Q^2}{R^2}\right )   \frac{r^2}{R^2} +  3 \frac{Q^2}{R^2}   \frac{r^4}{R^4}.
\end{equation} 
and

\begin{eqnarray}
\fl ~~~~~~~~~u(r) = \frac{1}{R^2} \int_0^r \frac{s ds}{\sqrt{  1-  \left ( \frac{2M}{R} +  \frac{2Q^2}{R^2} \right) \frac{s^2}{R^2}  +  \frac{3Q^2}{R^2} \frac{s^4}{R^4}   }}\\ \nonumber
\fl ~~~~~=  \frac{\sqrt{3} }{6}\frac{R}{ Q} \left( \log\left[ 1 + \sqrt{3}\left( \frac{M}{3Q} + \frac{Q}{3R}\right) \right] 
 - \log \left[  \sqrt{3}     \left( \frac{M}{3Q} + \frac{Q}{3R} - \frac{ Q r^2}{R^3}\right)         +         e^{-\lambda(r)}         \right] \right).
\end{eqnarray}

Thus the critical values  of $M/R$  vs $Q/R$ for $b  = 16 $ are solutions of the following equation:
\begin{equation}
\fl ~~~~~~~~\left( 1 - \frac{2 M}{R}  + \frac{Q^2}{R^2} \right)^{\frac{1}{2}}  = \left(\frac{M}{Q} - \frac{Q}{R} \right) \tan \left[ \frac{\sqrt{3}}{ 6}  \log \left[{\frac{ 1 +   \sqrt{3}       \left( \frac{M}{3Q} + \frac{Q}{3R}\right)} {   e^{-\lambda(R)}                     +  \sqrt{3}  \left( \frac{M}{3Q} - \frac{2Q}{3R} \right)}}\right] \right]
\end{equation} 

 \begin{figure}
\centering
 \includegraphics[width=0.75\linewidth]{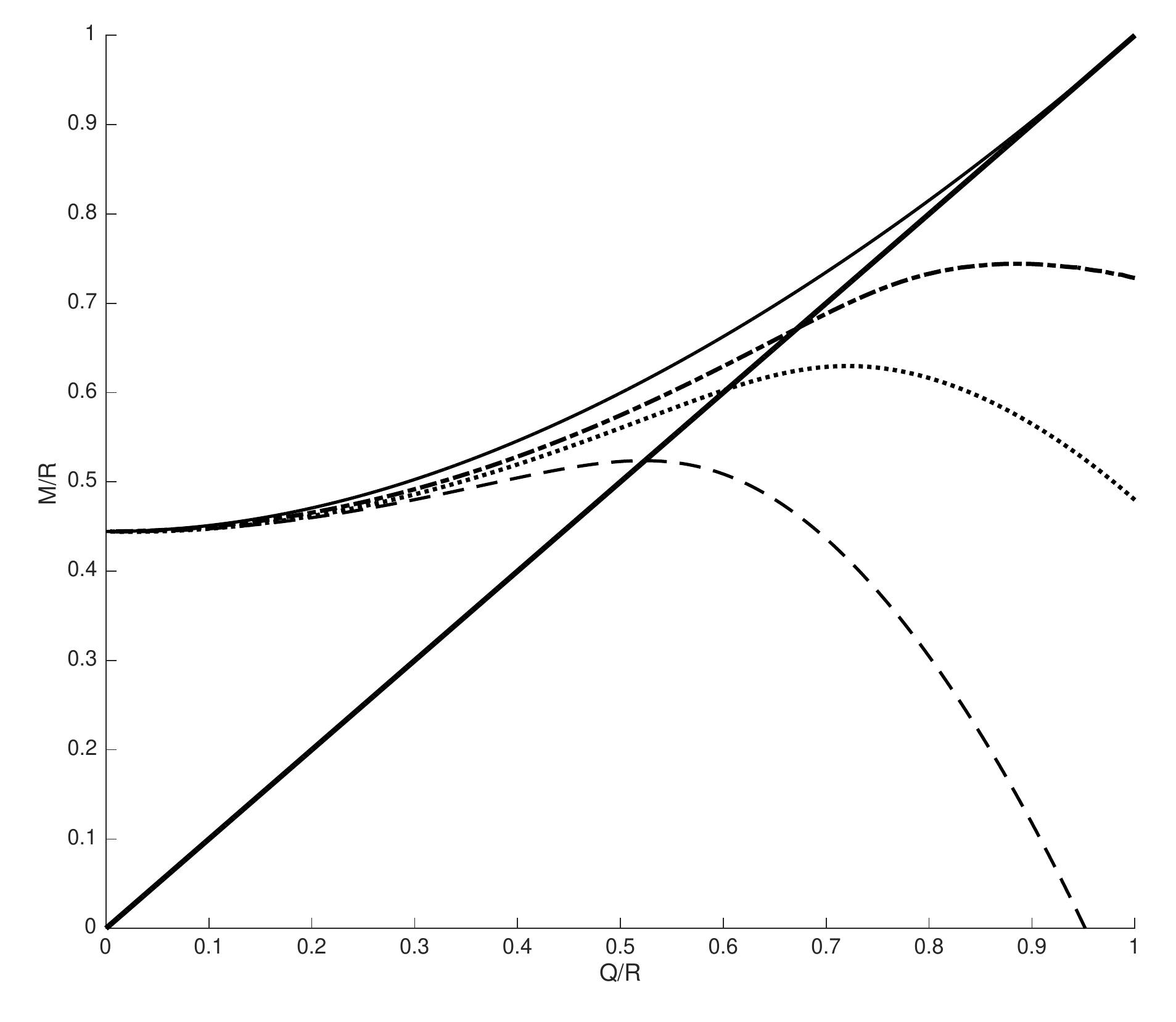}
  \caption{ \it  The critical values of M/R vs Q/R from the  $b = 31$ ({\bf- - - -}) ,  $b = 21$ ({\bf $\cdot$$\cdot$$\cdot$$\cdot$$\cdot$$\cdot$$\cdot$$\cdot$$\cdot$}), $b = 16$ ({\bf  $\cdot$ - $\cdot$ - $\cdot$ })  models and the Andr\'{e}asson   formula (\rule{0.9 cm}{0.03 cm}) are plotted.  The extremal line $M/R = Q/R$   (\rule{0.9 cm}{0.06 cm})             is also drawn.} 
\label{fig:f9}
\end{figure}

 \begin{table}[h!]
\centering
\begin{tabular}{|c| c |c| c|c|} 
 \hline
 b & $(\frac{M}{R})_{ext} $  & $ (\frac{Q}{R})_{max}$&  $ (\frac{Q}{R})_{peak}$   &$ (\frac{M}{R})_{peak}$\\ \hline
 
 16 & 0.66 & 1.0 & 0.92&0.744 \\ \hline 
 21 & 0.60 & 1.0 & 0.75& 0.63 \\ \hline
 26 & 0.558 & 1.0 & 0.68 & 0.564 \\ \hline
 31 & 0.523 & 0.952 & 0.56& 0.524 \\ \hline
 36 & 0.495 & 0.851 & 0.50&0.499 \\ \hline
 41& 0.471 & 0.766 &0.45&0.482 \\ \hline
 
\end{tabular}
\caption{ \it This table shows the change in various $\frac{M}{R}$ and $\frac{Q}{R} $ values with increasing $b$.  In column (ii)  $(M/R)_{ext}$ are the  extremal values  of $M/R$, in column (iii)   $(Q/R)_{max}$, gives the maximum value of $Q/R$ for which a non zero value of $M/R$ exists,   in column (iv)  the quantity $(Q/R)_{peak}$ is the value of $Q/R$ when the critical values curve attains its peak and  in column  (v)  $(M/R)_{peak}$ are the corresponding values $M/R$ at the peak.                                              }
\label{table:1}
\end{table}

\noindent  We solved this equation numerically and the values are plotted in Figure \ref{fig:f9}. The critical values of $M/R$ vs $Q/R$ for $b =  21 $ and $31$ are also plotted in Figure \ref{fig:f9}. We note that the  critical values  curves for these cases $( b > 11 )$ are distinctly  different from the  previous cases ($b \leq 11$) that we have already studied. The previous critical values curves essentially follow the shape of    the Andr\'{e}asson curve. In the current  case we  see an initial similarity with previous  curves,  however now the curves with $b > 11$ attains a peak before $Q/R  = 1$  and then decreases to a final value  of $M/R <1$ when $Q/R = 1$. These curves have extremal values of $M/R$ less than one.

Our  investigation of    cases with $ b \geq  11 $  found that both the peak values,  the extremal values  and the range of  values of $Q/R$ for which a non-zero value of $M/R$ exists  continuously decreases with increasing $b$ values.  In particular we found that when $  b \approx 29.25$ the limiting  value of $M/R$ is zero for $Q/R = 1$.  This result dictates  that a stable configuration with $b \approx 29.25 $ and $Q/R = 1$ does not exist. We also found that for $  b > 29.25 $ the maximum value of $Q/R$ for which a stable final configuration can exist decreases. These behaviors described here is shown in Figure \ref{fig:f9} and some values are given in Table 1.

\subsection{ Models with $b  < 0$}
  Andr\'{e}asson, in deriving  his  formula for the critical values of $M/R$ as a function of $Q/R$  for charged sphere, placed the following constraints on the matter content of the sphere: (i) $ p_r + 2 p_t < \rho$ and (ii) $\rho >0$. 

   Buchdahl  in his derivation of the critical value of $M/R$ for neutral perfect fluid spheres explicitly required that $d\rho/dr >0$.  Andr\'{e}asson  did not have such a requirement in the  derivation of his formula, in order to check the necessity of such a condition we studied charged spheres  with $ d\rho/dr > 0  $. 

The density profile for the spheres that we are studying is given by (\ref{rhoa}),  thus when $b < 0$,  $d\rho/dr >0$. We consider the case $  b = -9 $ in detail here and then comment on other solutions with $b < 0 $.

 \begin{figure}
\centering
 \includegraphics[width=0.75\linewidth]{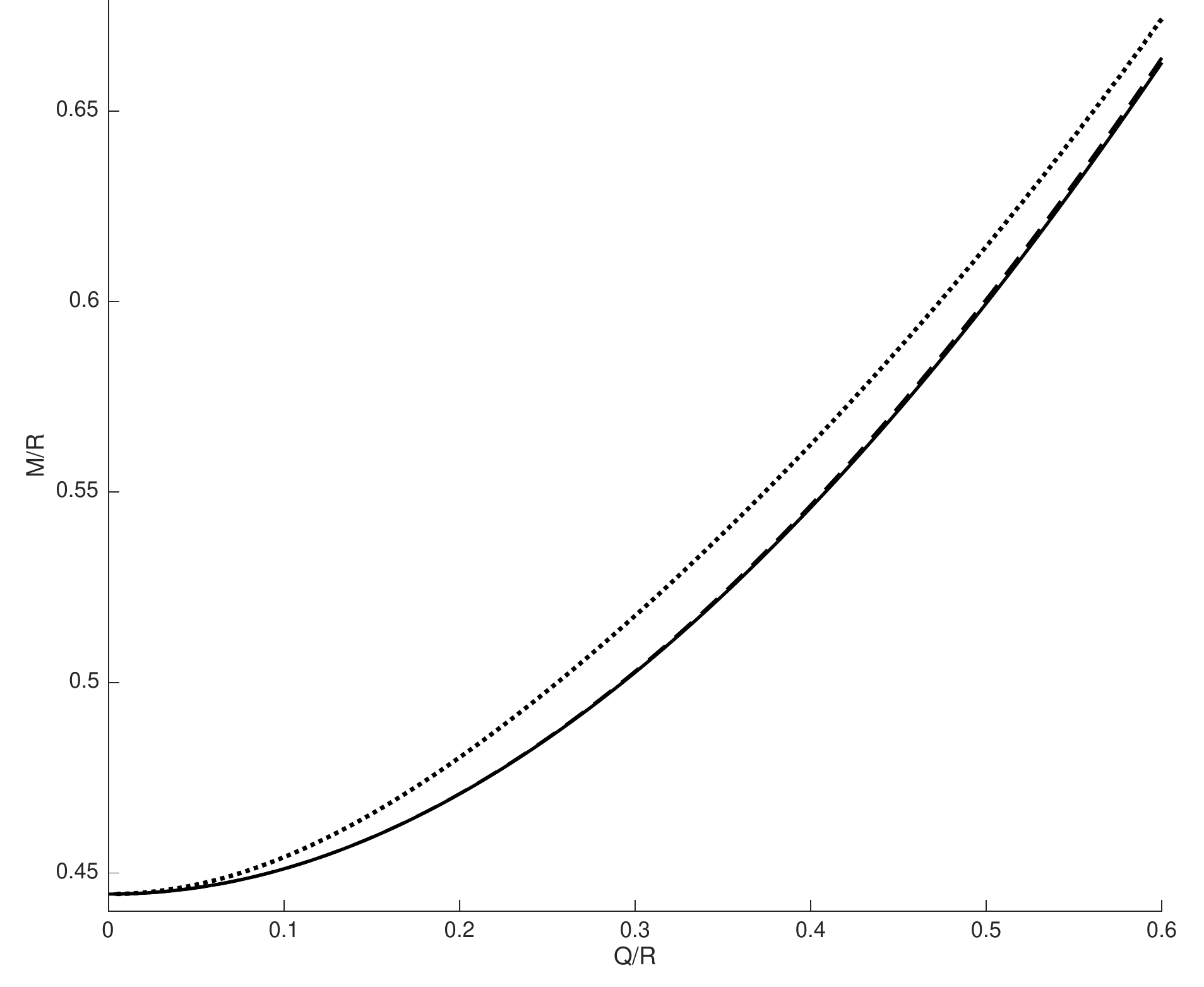}
  \caption{ \it The critical values of M/R vs Q/R  from the  $b = -69$  ({\bf ........}) and  $b =  -9$   $({\bf ----})$  models  and     the  Andr\'{e}asson formula   (\rule{0.9 cm}{0.03 cm}).      }
\label{fig:f10}
\end{figure}

When $b = -9$ 
\begin{equation}
\label{lambda}
e^{-2 \lambda(r)} = 1-  \left( \frac{2M}{R}  -3  \frac{Q^2}{R^2}    \right) \frac{r^2}{R^2}  -2  \frac{Q^2}{R^2} \frac{r^4}{R^4}.
\end{equation}
and
\begin{eqnarray}
\fl~~~~~~ u(r) = \frac{1}{R^2} \int_0^r \frac{s ds}{\sqrt{  1- \left( \frac{2M}{R}  - 3 \frac{Q^2}{R^2} \right) \frac{s^2}{R^2}  - 2 \frac{Q^2}{R^2} \frac{s^4}{R^4}   }}\\ \nonumber
\fl~~~~~~~~~~~ = \frac{{2}^{\frac{1}{2}}}{4}\frac{R}{Q} \left[ \arctan \left( {2}^{\frac{1}{2}} \left(\frac{M}{2Q}  - \frac{3}{4}\frac{Q}{ R }+ \frac{r^2Q}{R^3}\right) e^{\lambda(r)}
 \right)   -\arctan \left(  {2}^{\frac{1}{2}} \left(\frac{M}{2Q} - \frac{3Q}{4R}\right)\right) \right].
\end{eqnarray} 
The master equation here is 

\begin{equation}
\label{zeta1}
\frac{ d^2 \tilde{\zeta}}{d u^2}  =   a^2    \frac{Q^2}{R^2}  \tilde{\zeta} (u) {\rm ~~~with~~~} a^2 = \left(\frac{11}{5} - \frac{b}{5}\right)  
\end{equation}
The solution for this equation with $b = -9$  is
 \begin{equation}
 \tilde{\zeta} (u(r))  = A  e^{ 2\frac{Q}{R} u(r)  }           +  B e^{-2 \frac{Q}{R} u(r)  }.
 \end{equation}

Applying the boundary conditions we find
 \begin{equation}
A = \frac{1}{2} \left[ \sqrt{1 - \frac{2M}{R} + \frac{Q^2}{R^2}}  +    {\frac{1}{2}}   \left(\frac{M}{Q} - \frac{Q}{R}\right)\right] e^{-2 \frac{Q}{R} u(R)  },
\end{equation}
\begin{equation}
B = \frac{1}{2} \left[ \sqrt{1 - \frac{2M}{R} + \frac{Q^2}{R^2}} - {\frac{1}{2}} \left(\frac{M}{Q} - \frac{Q}{R}\right) \right] e^{2 \frac{Q}{R} u(R)  }.
\end{equation}
and the equation from which the critical values of $M/R$ vs $Q/R$ are obtained is :
\begin{eqnarray}
\fl  {\left[        \frac{1}{2}   \left(\frac{M}{Q} - \frac{Q}{R}\right)           +      \left(1 - \frac{2M}{R} + \frac{Q^2}{R^2}\right)^{\frac{1}{2}}               \right]} = { \left[        \frac{1}{2}     \left(\frac{M}{Q} - \frac{Q}{R}\right)           -      \left(1 - \frac{2M}{R} + \frac{Q^2}{R^2}\right)^{\frac{1}{2}}           \right] }   
  \\ \nonumber
\fl \times \exp \left [  \sqrt{2} \left(  \arctan\left(  \left( \frac{1}{\sqrt{2}}\frac{M}{Q}  + \frac{1}{2 \sqrt{2}}\frac{Q}{R}\right) e^{\lambda(R)}   \right)   - \arctan\left(\frac{1}{\sqrt{2}} \frac{M}{Q}  -     \frac{3}{2 \sqrt{2}} \frac{Q}{R}\right) \right) \right]
\end{eqnarray}

The critical values curve for $b = -9$ or $a = 2 $  is shown in Figure \ref{fig:f10}. We see that this model  saturates the  Andr\'{e}asson   limit. In our investigation of solutions for (\ref{zeta1}) we  found  that the increase in the critical values of $M/R$ with increasing negative $b$ values is relatively slow.  Therefore to show a noticeable  increase in the $M/R$ values we needed to plot a relatively large negative $b$ value. We  plotted the $ b = -69$ or $a = 4 $ curve in Figure \ref{fig:f10} and we observe that this model does  violate the  Andr\'{e}asson limit. Thus it is possible to exceed the upper limit for $M/R$ from  the  Andr\'{e}asson formula if $d\rho/dr > 0$. 

\section{Conclusion}
In this paper we studied exact solutions of the Einstein-Maxwell equations for the interior gravitational field of static spherically symmetric compact objects. We considered spheres that  consists of a perfect fluid with a charge distribution that gives rise to a radial static electric field. The inertial mass density of the fluid and the charged density associated with the electric field are given in (\ref{rhoa}). We explored in detail models with $b = 0, 1, 6, 11, 16  \, {\rm and} \, -9 $.  We computed the stability curves for these models and compared them with the stability curve from the 
Andr\'{e}asson formula. A summary of our  results are:
\begin{enumerate}
\item   All models with $b \geq 0$ have critical values of $M/R$ for a given $Q/R$ less than those predicted by the Andr\'{e}asson formula. 
\item  The critical values of $M/R$ for a given $Q/R$ decreases with increasing $b$ ($b > 0$). In particular we found that a sphere with a constant inertial mass density and constant charge density (the $b = 0$ model)  is more stable than a sphere with constant gravitational mass density and constant charge density (the $b = 6$ model).
\item  The critical values curves for models with $ 0 \leq b \leq 11 $  closely follows the contour of the Andr\'{e}asson curve. 
\item  The critical values curves for models with $  b  > 11$   initially follows the contour  of the Andr\'{e}asson curve but reaches a peak before $Q/R = 1$ and then turns sharply away from the 
Andr\'{e}asson curve. The value of $M/R$ when $Q/R = 1$ continuously decreases with increasing $b$ until $b \approx 29.25 $ then $M/R = 0$ for $Q/R = 1$.  For $b > 29.25$ the maximum value of $Q/R$ for which a stable configuration can exist continuously decreases.  
\item For $b > 11 $ extremal models can be formed with $Q/R = M/R \neq 1$.
\item For $ b < -9 $ the  critical values of  $M/R$ as function of $Q/R$ are larger than those predicted by the  Andr\'{e}asson formula, thus if the density of the fluid $\rho(r)$ has  $d\rho(r)/dr > 0$ it is possible to violate the   Andr\'{e}asson  limit. 
\end{enumerate}

\end{document}